\documentstyle[12pt,twoside,psfig]{article}

\def\be{\begin{equation}}
\def\ee{\end{equation}}
\def\bea{\begin{eqnarray}}
\def\ena{\end{eqnarray}}

\def\a{\alpha}
\def\b{\beta}

\def\d{\delta}
\def\g{\gamma}

\def\l{\lambda}
\def\m{\mu}
\def\n{\nu}

\def\q{\theta}                    %     \vartheta
\def\x{\xi}

\def\G{\Gamma}

\def\L{\Lambda}

\def\ch{{\cal H}}

\def\cl{{\cal L}}

\def\pa{\partial}
\def\xx{\times}
\def\dd{\bar{\d}}

\begin{document}

\thispagestyle{empty}
\begin{flushright}
{\sc ITP-SB}-95-43
\end{flushright}
\vspace{1cm}
\setcounter{footnote}{0}
\begin{center}
{\LARGE{On the renormalizability and unitarity of the Curci-Ferrari model
for massive vector bosons\footnote{This research was supported in part by NSF
grant no
 PHY9309888.}
    }}\\[14mm]

\sc{Jan de Boer\footnote{e-mail: deboer@insti.physics.sunysb.edu},
    Kostas Skenderis\footnote{e-mail: kostas@insti.physics.sunysb.edu},
    Peter van Nieuwenhuizen\footnote{e-mail:vannieu@insti.physics.sunysb.edu},
and Andrew Waldron\footnote{e-mail:wally@insti.physics.sunysb.edu},}\\[5mm]
{\it Institute for Theoretical Physics\\
State University of New York at Stony Brook\\
Stony Brook, NY 11794-3840, USA}\\[20mm]

{\sc Abstract}\\[2mm]
\end{center}
We prove the renormalizability of the Curci-Ferrari model with and
without auxiliary fields using BRST methods. In both cases
we find 5 $Z$ factors  instead of 3.
We verify our results by explicit one loop calculations.
We determine a set of generators for the ``physical states'', many
of which have negative norm.
Supersymmetrization is considered.

\vfill

\newpage

Recently, there has been renewed interest in consistent actions for
massive vector bosons without Higgs fields. In particular,
Periwal\cite{periwal} has reanalyzed the action of Curci and
Ferrari (\cite{c-f}, see also \cite{delb} and \cite{ojima})
for massive vector
bosons and claimed that it is both renormalizable and unitary. If true,
this would be astonishing since a great deal of work over the past 30 years
has left only the Higgs mechanism as a means of giving vector bosons
a mass while preserving renormalizability and unitarity.
Another recent suggestion is to use topological field theories\cite{niemi}.
In fact,
in his pioneering work in the 60's, Veltman\cite{veltman} started with massive
Yang-Mills theories coupled to free scalars, but after a field redefinition
which made the scalars seem to interact, the requirement of renormalizability
forced him to drop certain terms, thus ending up with the Higgs model.
Other studies\cite{others} which required that tree graphs do not grow too
fast with energy in order to obtain one-loop renormalizability confirmed
these results in the 70's.

A consistent model for massive vector bosons would not yet be an alternative
to Higgs fields since it is not immediately clear how to give
a gauge-invariant mass to the fermions of the standard model.
It would, however, allow infrared regularization of QCD and
of supersymmetric models (in particular in superspace)
both of which are plagued by serious infrared problems.

The model of Curci and Ferrari (CF) has
both a mass term for the vector bosons and a gauge fixing term. As a result the
propagator has  $k_\m k_\n /k^2$ terms instead of a non-renormalizable
$k_\m k_\n /m^2$ term, and is therefore power-counting renormalizable.
It also has a BRST (and an anti-BRST) symmetry, but here the analogy with
Higgs models stops: the BRST operator is neither nilpotent, nor can it be made
nilpotent by introducing a BRST auxiliary field. Despite much work in the
past, it is not clear whether this model can be obtained as a suitable
limit of a Higgs model. As a result of the non-nilpotency, the usual
``$\G \G$'' Ward identity of non-abelian gauge theories\cite{zinn} is modified
by an extra term, and the issue of renormalizability requires deeper
study. In this letter we first study the renormalizability of the CF model,
both with auxiliary field and without, and then we shall come back to the
issue of unitarity. We find that the theory is renormalizable,
as already found by Curci and Ferrari using different, more
cumbersome, methods, and by Periwal, but we find 5 $Z$ factors instead of 3,
as claimed by Periwal.
Then we determine the physical states, extending Ojima's work\cite{ojima}.
Many of these states have, for arbitrary values
of the parameters of the theory, a negative norm, and from this we
conclude that the model is not unitary. Finally we briefly discuss
supersymmetrization of the model.

The CF action is given by
\be
S = S_{\rm YM} + S_m + S_{\rm gf},
\ee
where $S_{\rm YM}$ is the Yang-Mills action
$-\frac{1}{4} (F_{\m \n}^a)^2$,
$S_m$ contains a mass for the vector bosons and the  ghosts $c^a$ and
antighosts $b^a$, while $S_{\rm gf}$ contains terms which resemble the
Faddeev-Popov ghost action and the gauge fixing term
\bea
S_m &=& -\frac{1}{2} m^2 ((A_\m^a)^2 + 2 \x b^a c^a), \\
S_{\rm gf} &=& \frac{1}{2} b^a (\pa^\m D_\m + D_\m \pa^\m) c^a
- \frac{1}{2\x} (\pa^\m A^a_\m)^2 + \frac{1}{8} g^2 \x (b \xx c)^2,
\ena
where $b \xx c \equiv f^a_{\ bc} b^b c^c$.
Both $S_m$ and $S_{\rm gf}$ are separately BRST invariant;
$A_\m^a$ and $c^a$ transform
as usual ($\d A_\m = D_\m c \L$ and $\d c = \frac{1}{2} g c \xx c \L$;
we omit the BRST parameter $\L$)
while the BRST law of $b^a$ can be found by requiring invariance
of $S_m$. It reads
$\d b = -\frac{1}{\x} \pa \cdot A + \frac{1}{2} g b \xx c$.
Since $\d \d b$ is non-vanishing, the action $S_{\rm gf}$ can be found
by assuming that for vanishing mass $\d \d b$ is proportional to the $b$ field
equation, and integrating the latter.

We first perform the analysis of renormalizability with a BRST auxiliary field
$\l^a$ present. If we define $\d b = \l$ but $\d \l = -m^2c$ (rather than
$\d \l = 0$ as nilpotency would require) the action $S-S_B$ is BRST invariant,
where
\bea
S &=&  S_{\rm YM} + S_m + S_{\rm gf}^\l + S_B, \label{action}\\
S_{\rm gf}^\l &=& S_{\rm gf} + \frac{1}{2} \x
(\l + \frac{1}{\x} \pa \cdot A - \frac{1}{2} g (b \xx c))^2 \\
&=& b^a \pa^\m D_\m c^a + \frac{1}{4} g^2 \x (b \xx c)^2
- A^a_\m \pa_\m \l^a
+ \frac{1}{2} \x \l^2 - \frac{1}{2} g \x \l \cdot (b \xx c) \nonumber \\
&=& \d [b^a (\frac{\x}{2} \l^a + \pa A^a - \frac{1}{4} g \xi (b \xx c)^a)]
+ \frac{1}{2} m^2 \xi b^a c^a, \nonumber \\
S_B &=& \int [ K \cdot DC + L\frac{1}{2}g c \xx c + M \cdot \l + N (-m^2 c)].
\ena
Following Zinn-Justin and B. Lee\cite{zinn}, we
couple the BRST variations to external sources.
We add source terms for the fields,
$S_s = \int [JA + lc + mb +n\l]$,
and obtain then the Ward identity for the effective action $\G = W - S_s$
\bea
&\ & (\frac{\pa}{\pa A} \G) \frac{\pa}{\pa K}\G
- (\frac{\pa}{\pa c} \G) \frac{\pa}{\pa L}\G
- (\frac{\pa}{\pa b} \G) \frac{\pa}{\pa M}\G
+ (\frac{\pa}{\pa \l} \G) \frac{\pa}{\pa N}\G \nonumber \\
&\ & \hspace{3cm}+ m^2 N \frac{\pa}{\pa L}\G
-M \frac{\pa}{\pa N}\G = 0.
\ena
As a check one may verify that $\G=S$ satisfies this equation.
In the renormalized theory, all $Z$ factors in this relation should
amount only to an overall rescaling. Hence, defining
$A_\m^a=Z_3^{1/2}A_\m^a({\rm ren})$ etc.,
$K_a^\m=Z_K^{1/2}K_a^\m ({\rm ren})$ etc., $\x = Z_\x \x({\rm ren})$,
$m^2 = Z_{m^2} m^2({\rm ren}), g = Z_g g({\rm ren})$,
we assume inductively the relations
\be
Z^{1/2}_3 Z^{1/2}_K = Z_c^{1/2} Z^{1/2}_L
= Z_b^{1/2} Z^{1/2}_M =
Z_\l^{1/2} Z^{1/2}_N = Z^{-1}_{m^2} Z^{-1/2}_N Z^{1/2}_L =
Z_M^{-1/2} Z^{1/2}_N. \label{scaling}
\ee
Since the action has vanishing ghost number, we expect that only the product
$(Z_b Z_c)^{1/2}=\tilde{Z}_3$ will be fixed.

Since $\G$ contains only one-particle irreducible graphs, it is independent of
$M$ and $N$, so we drop the term $-M \frac{\pa}{\pa N}\G$ term.
Assuming $(l-1)$ loop finiteness and the scaling hypothesis in (\ref{scaling}),
the $l$-loop divergences $\G({\rm div})$ must satisfy
$Q \G({\rm div}) = 0$, where
\be
Q= (\frac{\pa}{\pa x^i} S) \frac{\pa}{\pa \q_i}
+ (\frac{\pa}{\pa \q_i} S) \frac{\pa}{\pa x^i}
+m^2 N \frac{\pa}{\pa L}, \label{q}
\ee
with $x^i=\{ A, -L, -M, \l\}$ and $\q_i = \{K, c, b, N\}$.
We decompose $Q$ and $\G({\rm div})$ into terms without $m^2$ and terms
proportional to $m^2$
\bea
Q &=& Q^{(0)} + m^2 Q^{(1)}; \ \
\G({\rm div}) = \G^{(0)} + m^2 \G^{(1)}\\
Q^{(0)} &=& \frac{\pa}{\pa x^i} S(m^2=0) \frac{\pa}{\pa \q_i}
+ \frac{\pa}{\pa \q_i} S(m^2=0) \frac{\pa}{\pa x^i}, \\
Q^{(1)} &=&-A \frac{\pa}{\pa K} - c \frac{\pa}{\pa \l}
- \x b \frac{\pa}{\pa L} + \x c \frac{\pa}{\pa M}
\equiv \hat{Q}^{(1)} + \x c \frac{\pa}{\pa M}
\ena
where $S(m^2=0) = S_{\rm YM} + S_{\rm gf}^\l + K, L, M$ terms.
Again we drop the term $\x c \frac{\pa}{\pa M}$. We must then solve
\be
Q^{(0)} \G^{(0)} = 0; \ \ Q^{(0)} \G^{(1)} + \hat{Q}^{(1)} \G^{(0)} = 0;\ \
\hat{Q}^{(1)} \G^{(1)} = 0.
\ee
Since $Q^{(0)}$ is nilpotent (the term $N (-m^2 c)$ is no longer present
in $S(m^2=0)$) we have
\be
\G^{(0)} = \b S_{\rm YM} + Q^{(0)}[X+Y],
\ee
where $X$ and $Y$ denote the most general terms which are $M, N$ independent
and $M$ and/or $N$ dependent, respectively. Clearly,
\bea
X &=& \a_1 K \cdot A + \a_2 b \cdot \pa A + \a_3 L \cdot c
+\a_4 g b \cdot b \xx c + \a_5 b \cdot \l, \\
Y &=& \a_6 g N\cdot b \xx c + \a_7 g N\cdot N \xx c
+\a_8 N \cdot \l + \a_9 M \cdot b \nonumber \\
&\ & +\a_{10} M \cdot N + \a_{11} N \cdot \pa A.
\ena
$Q^{(0)} X$ is already $M$ and $N$ independent, but requiring
$Q^{(0)} Y$ to be $M, N$ independent  leaves only
\be
Y = \a_6 (M \cdot b + N \cdot \l).
\ee

The most general form of $\G^{(1)}$ is
\be
\G^{(1)} = \int (\g_1 A^2 + \g_2 \x b \cdot c),
\ee
which already satisfies $\hat{Q}^{(1)} \G^{(1)}=0$.
Hence we only need to solve $Q^{(0)} \G^{(1)} + \hat{Q}^{(1)} \G^{(0)} = 0$.
This yields the following three relations in the nine coefficients
$\b, \a_1, \ldots, \a_6, \g_1, \g_2$
\bea
&\ & 2\a_1 - \a_2 - \a_3 + \a_6 + 2\g_1 = 0 \\
&\ & 2\a_4 +g \x \a_6 + \frac{1}{2} g \x \g_2 = 0 \\
&\ & 2 \a_5 - 2 \x \a_6 - \x \g_2 = 0.
\ena
However, in $\G({\rm div})$ only 5 combinations of parameters occur
because the combination $Q^{(0)}(b \cdot N)$ in $X+Y$ is obviously
annihilated by $Q^{(0)}$.

The final form of the divergences is
\bea
\G({\rm div}) &=& \b S_{\rm YM} + \nonumber \\
&+& \a_1 [A\frac{\pa}{\pa A} S_{\rm YM} - m^2 A^2 + \l \cdot \pa A
- g \pa b \cdot A \xx c - K \cdot \pa c] \nonumber \\
&+& \a [-\frac{1}{2}  m^2 A^2 +  \l \cdot \pa A + b\pa^2c
- g \pa b \cdot A \xx c] \nonumber \\
&+& \g_2 [m^2 \x bc - \frac{1}{4} \x g^2 (b \xx c)^2 - \frac{1}{2} \x \l^2
+ \frac{1}{2} \x g \l \cdot b \xx c] \nonumber \\
&+& \a_3 [\frac{1}{2}  m^2 A^2 +\frac{1}{2} \x g^2 (b \xx c)^2
+b\pa^2c - g \pa b \cdot A \xx c \nonumber \\
&\ & \hspace{.5cm} -\frac{1}{2} \x g \l \cdot b \xx c + K \cdot \pa c
+ L\frac{1}{2} g c \xx c + K \cdot gA \xx c].
\ena
As a check we have verified that each of these terms is annihilated by $Q$.
In the process one needs the identities
\bea
&\ & (\frac{\pa}{\pa A} S_{\rm YM})\cdot \pa c =
\d (A\frac{\pa}{\pa A} S_{\rm YM}); \
(b \xx c)^2 = - \frac{1}{2}(b \xx b) \cdot (c \xx c); \\
&\ & (A \xx c) \xx c= \frac{1}{2} A \xx (c\xx c); \ \
c \cdot b \xx c = b \cdot c \xx c. \nonumber
\ena

Expanding the $Z$ factors as $Z_g=1-z_g, Z_3=1-z_A$ etc., we find
\bea
&\ &z_g=-\frac{1}{2} \b;\  z_A=2\a_1+\b;\ z_{m^2}=-\b - \a_3 +\a;\
z_\x = \b - \g_2 - 2\a \nonumber \\
&\ &z_\l = 2 \a - \b; \
\frac{1}{2}(z_b +z_c)=\a +\a_3;\ (z_K+z_c)=2\a_3-2\a_1; \nonumber\\
&\ & \frac{1}{2}z_L+z_c=\a_3+\frac{1}{2}\b.
\ena
As expected from ghost number conservation, only $z_b +z_c$, $z_K+z_c$
and $z_L+2z_c$ occur. Since $M$ and $N$ do not contribute to
$\G$ except at the tree level, we further get
$z_M=-z_\l$ and $z_N=-2z_{m^2}-z_c$.
One can easily check that these $Z$ factors satisfy the scaling
hypothesis (\ref{scaling}). This completes the proof by induction
of the renormalizability of the model with the auxiliary field present.
There are clearly 5 $Z$ factors.

The action has an infinitesimal $U(1)$ invariance $\d c=b, \d b=-c,
\d \l=\frac{1}{2} g b \xx b - \frac{1}{2} g c \xx c$,
leading to a finite symmetry
\be
c \rightarrow b, b \rightarrow -c, \l \rightarrow \l -g b \xx c.
\ee
Under this symmetry $\cl_m$, $\cl_{{\rm gf}}$ and
$(\l + \frac{1}{\x} \pa \cdot A - \frac{1}{2} g (b \xx c))$
are separately invariant.
However, $\G({\rm div})$ does {\em not} have this symmetry even at
$K=L=0$, nor should it, since $\d\l$ is nonlinear in fields and hence is
modified at the quantum level. Since $S+\G({\rm div})$ can be written
as the renormalized $S$, it is clear that the renormalized  transformation rule
for $\d\l$ keeps $S+\G({\rm div})$ $U(1)$ invariant.
Requiring (erroneously) that only $\G({\rm div})$ be $U(1)$ invariant would
lead to $\a_3=0$, but we keep $\a_3$.

Eliminating $\l$ by substituting the $\l$ field equation
$\l=-\frac{1}{\x} \pa \cdot A + \frac{1}{2} g  (b \xx c)-\frac{1}{\x}M$
in $S$ and in $\G({\rm div})$ we find
\bea
S \!\!\!&=&\!\!\!  S_{\rm YM} + S_m + S_{\rm gf} \nonumber\\
&&\!\!\!+ \int[ K \cdot DC + L\frac{1}{2}g c \xx c +
M(-\frac{1}{\x} \pa \cdot A + \frac{1}{2} g  (b \xx c))
-\frac{1}{2 \x} M^2],\label{elvis}
\ena
and
\bea
\G({\rm div, no} \hspace{.2cm} \l) &=& \b S_{\rm YM} + \nonumber \\
&+& \a_1 [A\frac{\pa}{\pa A} S_{\rm YM} - m^2 A^2 -\frac{1}{\x}(\pa A)^2
+\frac{1}{2}g A_\m j^\m- K \cdot \pa c -\frac{1}{\x}M \cdot \pa A] \nonumber \\
&+& \a [-\frac{1}{2}  m^2 A^2 -\frac{1}{\x}(\pa A)^2 + b \pa^2 c
+\frac{1}{2}gA_\m j^\m - \frac{1}{\x} M \cdot \pa A] \nonumber \\
&+& \g_2 [m^2 \x bc - \frac{1}{8} \x g^2 (b \xx c)^2 - \frac{1}{2\x}(\pa A)^2
-\frac{1}{\x} M \cdot \pa A-\frac{1}{2 \x} M^2] \nonumber \\
&+& \a_3 [\frac{1}{2}  m^2 A^2 +\frac{1}{4} \x g^2 (b \xx c)^2
+b \pa^2 c+\frac{1}{2}gA_\m j^\m  \nonumber \\
&\ & + K \cdot \pa c + L\frac{1}{2} g c \xx c + K \cdot gA \xx c
+\frac{1}{2}g M \cdot b \xx c],\label{presley}
\ena
where $j^\m=\pa^\m b \xx c - b \xx \pa^\m c$. Notice that $A$ only couples to
the ghosts through the $U(1)$ current. All the terms in
$\G({\rm div, no}\ \l)$
are $U(1)$ invariant when $K=L=M=0$.
One can easily check that the same $Z$-factors still render the model
finite.

Furthermore, as an additional check of our results, we began with the
action (\ref{elvis}) without auxiliary fields and performed an analysis
similar to that given above. This is possible because the term
$-(1/2\xi )M^2$ plays the r\^{o}le of the usual subtraction
$S\rightarrow S-S_{\mbox{fix}}$ that one makes when solving the BRST cohomology
in the usual case of non-abelian gauge theories\cite{zinn}.
Of course, the results are exactly as given in (\ref{presley}).

Periwal in \cite{periwal} has shown that the same model is renormalizable,
however with two fewer parameters as we now discuss.
Comparing our results to those of Periwal, we find that his results
are a subcase of ours in the following sense
\be
Z_\x=\tilde{Z}_3^{-3}Z^{-1/2}_3Z_g^{-3};
Z_{m^2} =\tilde{Z}_3^{3}Z_3Z_g^{4}. \label{p1}
\ee
The conditions in (\ref{p1}) imply
\be
\g_2=\a_3;\hspace{1cm} \a+\a_1+2\a_3=0 \label{conditions}
\ee
Then he finds
\be
Z_K=\tilde{Z}_3^{-3}Z^{-2}_3Z_g^{-4};
Z_L=Z_M=\tilde{Z}_3^{-4}Z^{-1}_3Z_g^{-4}  \label{p2}
\ee
which agrees with our results if we use (\ref{conditions}).
To check that the relations in (\ref{conditions}) are not due to a symmetry
which we overlooked, we made an analysis of divergences at the one
loop level.

\begin{table}
\begin{equation}
\begin{array}{ll}
g^2 C_2 D  \int b \partial^2 c \left(-\frac{3}{4} + \frac{\xi}{4}\right)
 \hspace{5mm}
   & \alpha+\alpha_3 \\
g^2 C_2 D  \int b\xi m^2 c \left(\frac{-\xi}{4}    \right)
 \hspace{5mm}
& \gamma_2 \\
g^2 C_2 D  \int K \partial c \left( \frac{-3}{4}   \right)
 \hspace{5mm}
 & \alpha_3-\alpha_1 \\
g^2 C_2 D  \int g K \cdot A \times c \left( \frac{\xi}{4}   \right)
 \hspace{5mm}
 & \alpha_3 \\
g^2 C_2 D  \int \frac{1}{2} g L \cdot c \times c \left( \frac{\xi}{4}
      \right)
 \hspace{5mm}
 & \alpha_3 \\
g^2 C_2 D  \int \frac{1}{2} g A \cdot (\partial b \times c -
 b \times \partial c) \left( \frac{\xi}{2}   \right)
 \hspace{5mm}
 & \alpha + \alpha_1 + \alpha_3 \\
g^2 C_2 D  \int m^2 A^2 \left( - \frac{3}{8}-\frac{\xi}{8}   \right)
 \hspace{5mm}
 & -\alpha_1 - \frac{\alpha}{2} + \frac{\alpha_3}{2} \\
g^2 C_2 D  \int (\partial_{\mu} A_{\nu})^2   \left(
      \frac{13}{12}-\frac{\xi}{4}   \right)
 \hspace{5mm}
 & -\frac{\beta}{2} -\alpha_1 \\
g^2 C_2 D  \int (\partial \cdot A)^2 \left( \frac{-29}{24}
      +\frac{\xi}{4}   \right)
 \hspace{5mm}
 & \frac{\beta}{2} + \alpha_1 - \frac{\alpha_1}{\xi}
   -\frac{\alpha}{\xi} -\frac{\gamma_2}{2\xi} \\
g^2 C_2 D  \int \frac{-1}{\xi} M \cdot \partial A \left(  0  \right)
 \hspace{5mm}
  & \alpha + \alpha_1 + \gamma_2 \\
g^2 C_2 D  \int \frac{-1}{2\xi} M^2 \left( \frac{-\xi}{4}   \right)
 \hspace{5mm}
  & \gamma_2 \\
g^2 C_2 D  \int \frac{1}{2} g M\cdot b \times c \left( \frac{\xi}{4}
 \right) & \alpha_3
\end{array}
\end{equation}
\caption{One loop divergences. The column on the right indicates
which parameters are fixed by the corresponding divergence.}
\end{table}
Table 1 gives the result of one loop calculations for the divergences
indicated, followed by combinations of parameters which get fixed in this way.
The number $C_2$ is defined by $f_{ap}^{\ \ q} f^{b\ p}_{\ q} = -C_2 \d_a^b$,
and $D$ denotes the standard divergence
$\int d^4k (2\pi)^{-4} (k^2 + m^2)^{-2}$. From these results
we conclude that at the one loop level the divergences
are given, up to an overall factor $g^2 C_2 D$ by
\be
\b = - \frac{11}{3}, \ \ \a_1 = \frac{1}{4} (3+\x), \ \
\a= - \frac{3}{4}, \ \ \g_2 = - \frac{1}{4} \x, \ \ \a_3 = \frac{1}{4} \x
\label{result}
\ee
Clearly, the conditions in (\ref{conditions})  disagree with the values in
(\ref{result}). In particular the quantity $g A/m^2 \x$ does renormalize.

The simplest check that the relations (\ref{conditions}) do not hold,
is given by the graphs in figure 1. The first one yields $\g_2$,
while the sum of the last two graphs yield $\a_3$. Clearly, at the one loop
level
\be
\a_3 = - \g_2,
\ee
which does not agree with (\ref{conditions}).
\begin{figure}
$$
\begin{array}{ccc}
\hbox{\psfig{figure=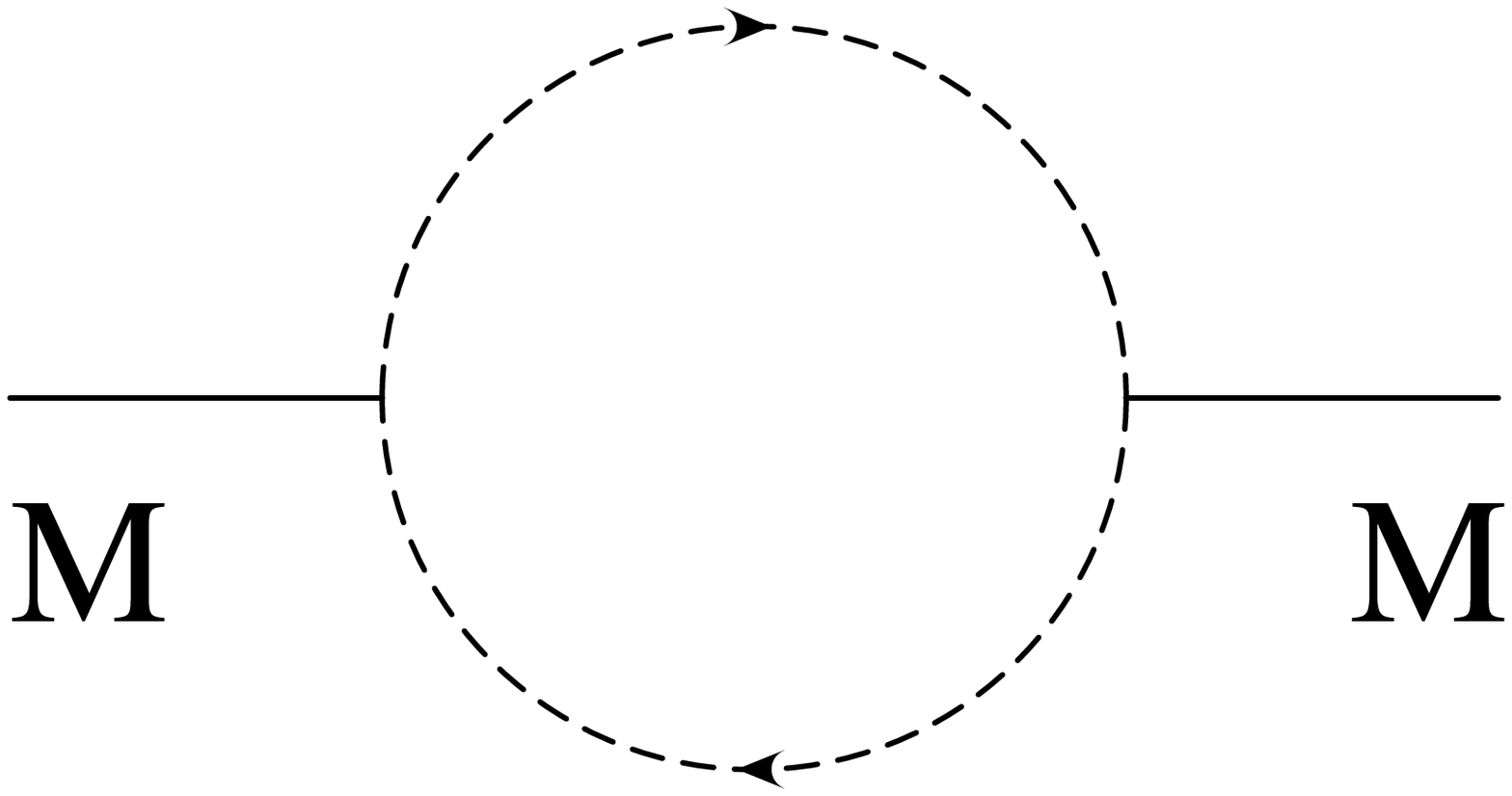,width=4cm}}&
\raisebox{-.6cm}{\psfig{figure=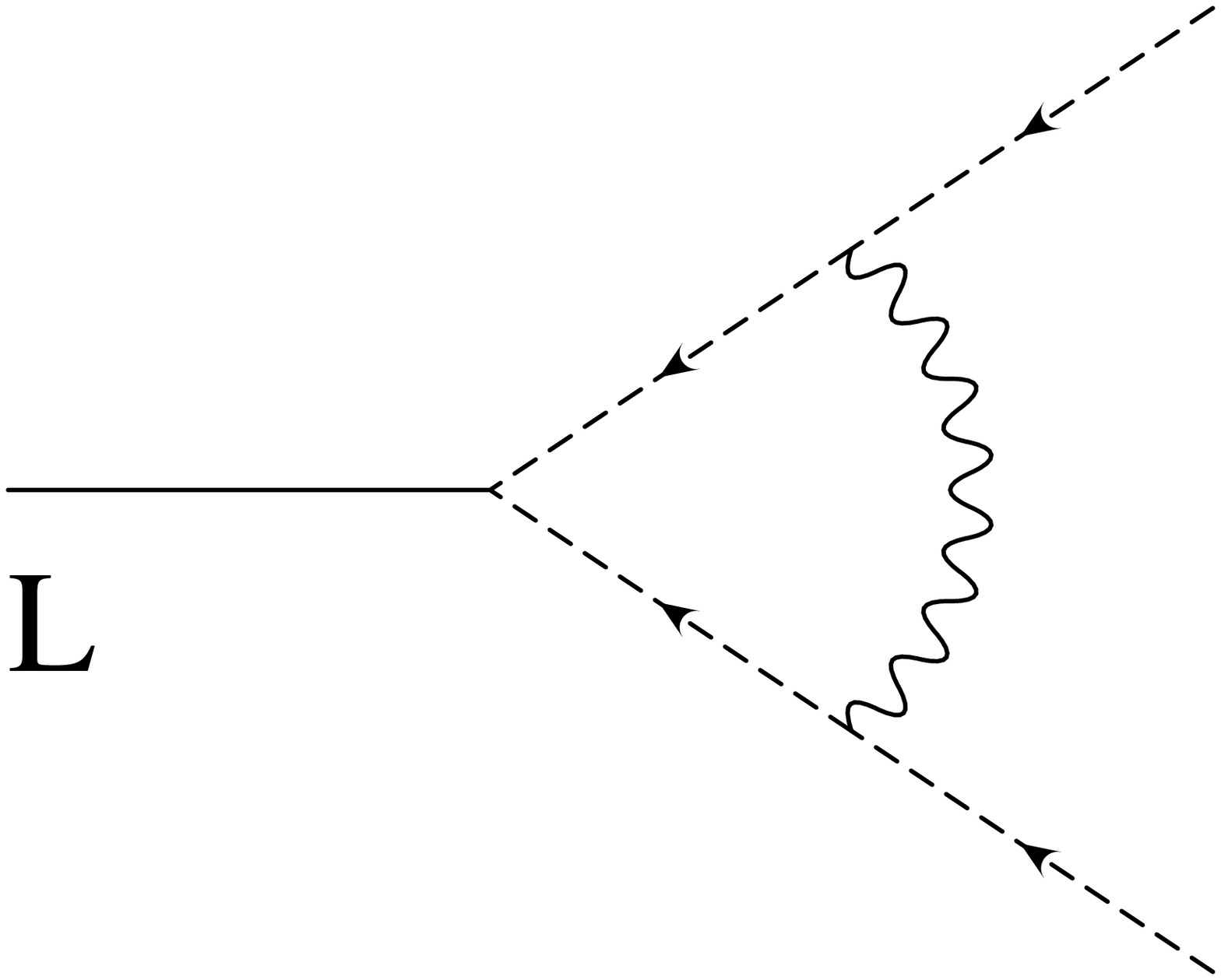,width=4cm}}&
\hbox{\psfig{figure=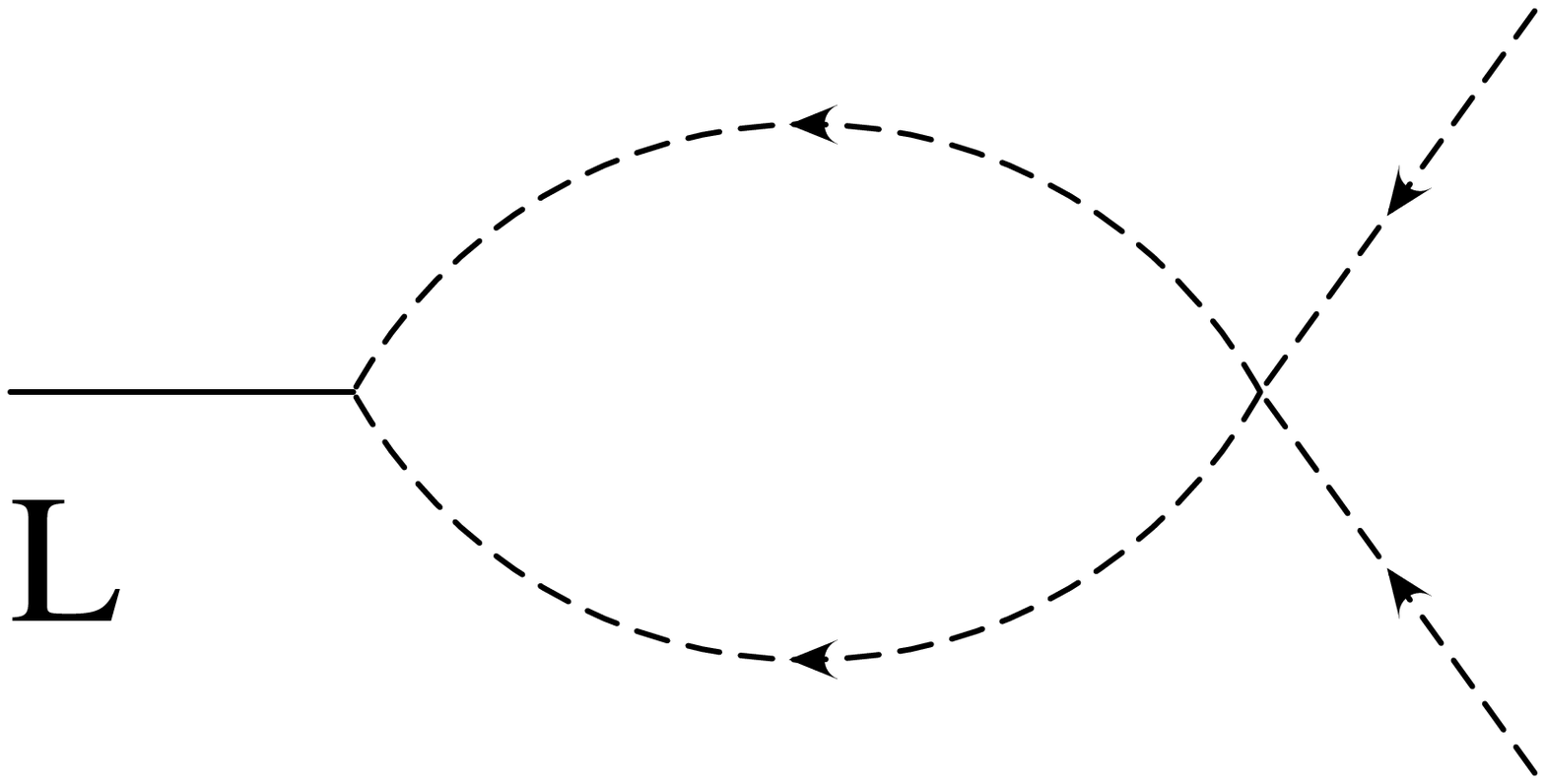,width=4cm}}\\
-\frac{1}{4}g^2\xi C_2 D &\frac{1}{8}g^2\xi C_2 D&
\frac{1}{8}g^2\xi C_2 D \nonumber
\end{array}
$$
Figure 1.
\end{figure}

Although the model is renormalizable, it does not seem to be unitary.
We perform a Hamiltonian analysis to determine the spectrum of the
theory. We assume the usual relation between Heisenberg fields and in- and
out- states\cite{k-o}. We eliminate $A_0^a$ and $p(A_0^a)$ from the
Hamiltonian by using the second class constraints $p(\l) - A_0 =0$
and $p(A_0)=0$.
A canonical transformation with generator
\be
F_2(q, P)=\int d^3x[(A_i^a + \frac{1}{m^2} \pa_i \l^a) P^i_a(U)
+\l^a P_a(\L)]
\ee
decouples the $A$-terms from the $\l$- terms in the Hamiltonian.
The Hamiltonian density in terms of the new asymptotic fields
$U_i^a$ and $\L^a$ reads
\bea
\ch &=& \frac{1}{2} P^i (\d_{ij} - \frac{\pa_i \pa_j}{m^2}) P^j
- \frac{1}{2} U_i (\pa^2 \d_{ij} - \pa_i \pa_j - m^2 \d_{ij}) U_j \nonumber \\
&\ & - \frac{1}{2} m^2 P^2 + \frac{1}{2 m^2} \L (\pa^2 - m^2 \x)\L
-p(c) p(b) - b(\pa^2 - m^2 \x)c,
\ena
where $\pa^2 = \pa^i \pa_i$.
Note the unphysical sign in front of the the $\L$ terms.
All field equations are of Klein-Gordon type with $m^2$ for $U_i$ and
$P_i$, and $\x m^2$ for $\L$ and $P(\L)$.

The corresponding modes satisfy the commutation relations
\be
[a^a_i(\vec{k}), a^{b}_j(\vec{l})^\dagger] =
\d^{ab} (\d_{ij} + \frac{k_i k_j}{m^2}) \d(\vec{k} - \vec{l}),
\ee
\be
[a_\L^a (\vec{k}), a_\L^b (\vec{l})^\dagger] =
 - \d^{ab} \d(\vec{k} - \vec{l}).
\ee
For the ghost/antighost system one finds the usual
$\{ c^a(\vec{k}), b^b(\vec{l})^\dagger \} = \d^{ab} \d(\vec{k} - \vec{l})$.
The BRST tranformations read
\be
\d a_i(\vec{k}) = 0; \ \d b(\vec{k}) = m a_\L(\vec{k}); \
\d a_\L(\vec{k}) = m c(\vec{k}); \ \d c(\vec{k}) = 0.
\ee
These results are the canonical counterpart of Ojima's analysis in
\cite{ojima}, where he shifted $A_\m^a$ into
$U^a_\m = A_\m^a + \pa_\m \l^a / m^2$, after which $U^a_\m$ and $\l^a$
decouple in configuration space.

Defining ``physical states'' to be BRST and anti-BRST invariant, all
$a_i(\vec{k})^\dagger$ generate physical states. However, in contrast to the
usual case, there are further physical states made up from ghosts,
antighosts and $\L$ oscillators, some of which have negative
(and $\xi$-independent!) norm. One might expect this since $Q$ is not
nilpotent, as a result of which the unphysical states do not form
Kugo-Ojima quartets. We restrict our analysis here to the case with
the auxiliary field present, but similar remarks apply also to the formulation
without the auxiliary field. In that case there is a particularly simple
ghost-dependent observable, namely the mass-term in the action.

To find the states in the formulation with the auxiliary field,
we note that the anti-BRST
transformations read (from now on we take $m=1$)
\be
\dd a_i(\vec{k}) = 0;\ \dd c(\vec{k}) = - a_\L(\vec{k}); \
\dd a_\L(\vec{k}) = b(\vec{k}); \dd b(\vec{k}) = 0.
\ee
Clearly, $\d, \dd$ and $\{\d, \dd \}$ form a graded version of $sl(2)$,
and $(b, \l, c)$ form a spin-1 representation. The BRST and anti-BRST
invariant states are the singlets of this graded $sl(2)$.
To find the physical states we must determine the singlets in tensor products
of the spin-1 representations. This problem has been solved for the usual
(ungraded) $sl(2)$ case by Weyl\cite{weyl}.

All invariants are built either from inner products $(\vec{v}_1,\vec{v_2})$,
or from determinants $\det( \vec{v}_1, \vec{v}_2, \vec{v}_3)$, with
$\vec{v}_i$ three-component spin-one vectors. Adapted to our case, this
yields the following (overcomplete) set of generators of physical states
\bea
A^r(k_1) & = & (r+2)\lambda_1^r b_1 c_1 - \lambda_1^{r+2} \\
A^{r_1,r_2}(k_1,k_2) & = &
H_1 H_2 -\lambda_1^{r_1} \lambda_2^{r_2} (b_1 c_2 + b_2 c_1) \\
A^{r_1,r_2,r_3}(k_1,k_2,k_3)  & = &
H_1 \lambda_2^{r_2} \lambda_3^{r_3} (b_2 c_3 + b_3 c_2) \nonumber \\
& & +
H_2 \lambda_3^{r_3} \lambda_1^{r_1} (b_3 c_1 + b_1 c_3) \nonumber \\
& & +
H_3 \lambda_1^{r_1} \lambda_2^{r_2} (b_1 c_2 + b_2 c_1) \nonumber \\
& & -2H_1 H_2 H_3
\ena
where
\be H_i=\lambda_i^{r_i+1} - r_i \lambda_i^{r_i-1} b_i c_i, \ee
the $r_i$ are non-negative integers, $k_1 \neq k_2 \neq k_3$,
and $\lambda_i=a_\L^{a_i}(k_i)^{\dagger}$,
$b_i=b^{a_i}(k_i)^{\dagger}$ and $c_i=c^{a_i}(k_i)^{\dagger}$.
To obtain a physical state, we still need to contract a product
of these generators with a suitable tensor, so as to make the
final result invariant under global gauge transformations. Under the
latter, $A^r(k_1)$ transforms as $(T^{a_1})^{r+2}$, $A^{r_1,r_2}(k_1,
k_2)$ as $(T^{a_1})^{r_1+1} (T^{a_2})^{r_2+1}$, etc.
None of the states made up out of these generators are of the form
$\delta X + \bar{\delta} Y$. They have vanishing ghost number and
are invariant under the linearized
$U(1)$ symmetry $b\rightarrow c$, $c\rightarrow -b$.
The state $A^{0,0}(k,l)|0\rangle$ was found by Ojima \cite{ojima},
who observed it has a negative norm. In fact, many of the physical states
in this sector of the Hilbert space have negative, $\xi$-independent norm,
as can be seen from the examples
\bea
 | A^r(k) |0\rangle |^2  &  =  & (-1)^{r+1}(r+2)r! \\
 | A^{r_1,r_2}(k_1,k_2) |0\rangle |^2  &  =  & (-1)^{r_1+r_2+1} r_1!
   r_2! \\
 | A^{r_1,r_2,r_3}(k_1,k_2,k_3) |0\rangle |^2  &  =  & 2 (-1)^{r_1+r_2+r_3}
    r_1! r_2! r_3!
\ena

We can conclude that the CF model is renormalizable but not unitary
for any value of $\x$. Of course, this argument would break down if
the relation between Heisenberg fields and in- and out- states
\cite{k-o} is no longer valid in this model. This might for instance
happen if the theory only makes sense if it is strongly coupled,
or has bound states, but then it is not clear to what extent a
perturbative analysis can be trusted.

%It can presumably be coupled to matter because
%the usual problems due to $k_\m k_\n /m^2$ terms in the propagator as
%studied in ref. \cite{others} are absent here.

Although the model is not unitary, it might be useful as a regularization
scheme for infrared divergences. In particular, in superspace,
where dimensional regularization is incompatible with supersymmetry,
this scheme may finally resolve long-standing problems concerning
infrared divergences. To supersymmetrize the model, one might
start from the observation that the double BRST variation of the antighost is
proportional to the antighost field equation for vanishing mass.
Alternatively one might seek an action of the form
$S_{gf}=Q^{(0)} \int d^{8}z \mbox{Tr} b [\bar{\l} + \bar{F}]+\mbox{h.c.}$
and try to solve $Q^{(1)}S_{gf}+Q^{(0)}S_m=0$ for $\bar{F}$.
In the former case one finds the BRST transformation rule
for the chiral antighost $b$ from the invariance of a mass term
$S_{m}= (m^2/g^2)\int d^8 z\mbox{Tr}e^V
+ (m/g^2) (\int d^6z \mbox{Tr}bc + \mbox{h.c.})$
using the BRST transformations $\delta e^V=-\bar{c} e^V +e^V c$,
$\delta c=-c^2$ (here $\d$ is an antiderivation). The result is
$\delta b=- m \bar{D}^2 e^V - (1/2)\{ b, c\}$.
One can then write a BRST invariant action
\begin{eqnarray}
S&=&S_{\rm {SYM}}+S_{m}
+\frac{1}{g^2\xi} \left[
\int d^8z (-b(-\bar{c} e^V+e^V c)-\frac{1}{2} \{ b, c\} e^V) \right.
\nonumber \\
&+&\left. \int d^6z(\frac{1}{8m}\{ b, c\}^2-\frac{m}{2}(\bar{D}^2 e^V)^2)+h.c.
\right]
\end{eqnarray}
We note that this action does not have a gauge fixing of the form
$(\partial \cdot A)^2$ in the bosonic sector. Further work on the
supersymmetrization of the CF model and its applications is in
progress\cite{us}.


\begin{thebibliography}{99}
\bibitem{periwal} V. Periwal, {\it Infrared regularization of non-Abelian
gauge theories}, PUPT-1562 {\tt hep-th/9509084}; {\it Unitary theory of
massive non-Abelian vector bosons}, PUPT-1563 {\tt hep-th/9509085},
to appear.
\bibitem{c-f} G. Curci and R. Ferrari, Nuo. Cim. {\bf 32A} (1976) 151;
{\bf 35A} (1976) 1.
\bibitem{delb} R. Delbourgo, S. Twisk and G. Thompson, Int. J. Mod. Phys.
{\bf A3} (1988) 435;
G. Curci and E. d'Emilio, Phys. Lett. {\bf 83B} (1979) 199.
\bibitem{ojima} I. Ojima, Z. Phys. {\bf C13} (1982) 173.
\bibitem{niemi} A.J. Niemi, {\it Gauge vector masses from flat connections},
UU-ITP-15/95, {\tt hep-ph/9510201}.
\bibitem{veltman} M. J. Veltman, Nucl. Phys. {\bf B7} (1968) 637.
\bibitem{others} C. H. Llewellyn Smith, Phys. Lett. {\bf 46B} (1973) 233;
J. M. Cornwall, P. V. Levin and G. Tiktopoulos, Phys. Rev. Lett. {\bf 30}
(1973) 1268, Phys. Rev. {\bf D10} (1974) 1145.
\bibitem{zinn} J. Zinn-Justin, in ``{\it Trends in Elementary Particle
Theory}'', Lecture notes in Physics vol. 37, Springer Verlag, Berlin, 1975,
eds. H. Rollnik and K. Dietz; B. Lee, in ``{\it Methods in Field Theory,
Proceedings Les Houches session 28}'', 1975, eds. R. Balian and
J. Zinn-Justin, North-Holland, Amsterdam, 1976.
\bibitem{k-o} T. Kugo, I. Ojima, Phys. Lett. {\bf 83B} (1979) 93;
Prog. Theor. Phys. {\bf 60} (1978) 1869, {\bf 61}, (1979) 294, 644;
Prog. Theor. Phys. Suppl. {\bf 66} (1979) 1.
\bibitem{weyl} H. Weyl, {\it The classical groups, their invariants
and representations}, Princeton, New Jersey, Princeton University Press,
1946.
\bibitem{us} J. de Boer, M. Ro\v{c}ek, K. Skenderis, P. van Nieuwenhuizen
and A. Waldron, in preparation.
\end{thebibliography}
\end{document}